# Up-sampling and Natural Sample Value Computation for Digital Pulse Width Modulators

Kien C. Nguyen
Department of Electrical and Computer Engineering
and the Coordinated Science Laboratory
University of Illinois at Urbana-Champaign
Urbana, IL 61801
Email: knguyen4@uiuc.edu

Dilip V. Sarwate
Department of Electrical and Computer Engineering
and the Coordinated Science Laboratory
University of Illinois at Urbana-Champaign
Urbana, IL 61801
Email: sarwate@uiuc.edu

*Abstract*— Digital pulse width modulation has been considered for high-fidelity and high-efficiency audio amplifiers for several years. It has been shown that the distortion can be reduced and the implementation of the system can be simplified if the switching frequency is much higher than the Nyquist rate of the modulating waveform. Hence, the input digital source is normally upsampled to a higher frequency. It was also proved that converting uniform samples to natural samples will decrease the harmonic distortion. Thus, in this paper, we examine a new approach that combines upsampling, digital interpolation and natural sampling conversion. This approach uses poly-phase implementation of the digital interpolation filter and digital differentiators. We will show that the structure consists of an FIR-type linear stage and a nonlinear stage. Some spectral simulation results of a pulse width modulation system based on this approach will also be presented. Finally, we will discuss the improvement of the new approach over old algorithms.

*Type: Regular paper*

## I. INTRODUCTION

### A. PWM Signals

A *pulse width modulation* (PWM) signal consists of a sequence of pulses of varying width or duration. One pulse occurs every $T$ seconds and its width (in the range $(0, T)$) is proportional to a sample value of the modulating signal $x(t)$ where, without loss of generality, we assume that $|x(t)| < 1$. In this paper, we consider only *trailing-edge* PWM signals in which the leading edges occur at fixed times $kT$ while the position of the trailing edge varies. In the first graph in Fig. 1, we illustrate a modulating signal $x(t)$ and a saw-tooth waveform (carrier ramp) of frequency $f_c = T^{-1}$. The second graph in Fig. 1 illustrates a *uniform-sampling* PWM signal in which the sample value $x(kT)$ determines the width of the $k$-th pulse. Note that the samples of $x(t)$ are *uniformly spaced in time*. The third graph of Fig. 1 illustrates a *natural-sampling* PWM signal in which the pulse width is determined by the intersection of $x(t)$ and the carrier ramp.[1] Thus, the sample values that determine the pulse widths are generally not uniformly spaced in time. Natural sampling seemingly appears to be more complicated than uniform sampling, but is in fact relatively easy to implement. Applying $x(t)$ and the carrier ramp to the inputs of an operational amplifier without feedback produces the natural sampling PWM signal.

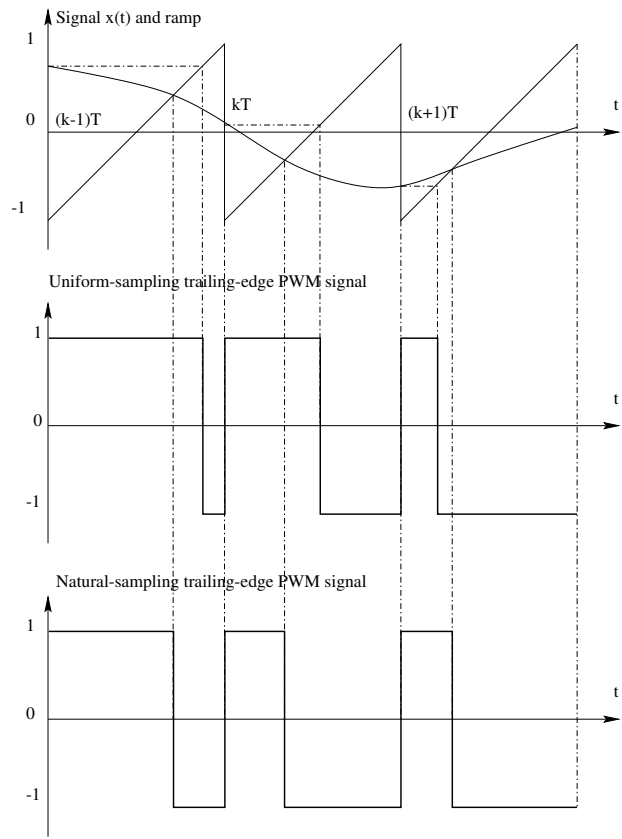

Fig. 1. Uniform-sampling and natural-sampling trailing-edge PWM signals

### B. Modulation and Demodulation

Switching amplifiers can be used to amplify PWM signals with power efficiencies far exceeding the theoretical 50% efficiency achievable by linear amplifiers. Furthermore, a simple demodulator for natural-sampling PWM signals is a low-pass filter with cut-off frequency $f_L$ satisfying $f_s < f_L \ll f_c$. An ideal low-pass filter produces $x(t)$ exactly without *any* harmonic distortion. In contrast, low-pass filtering of uniform-sampling PWM signals recovers $x(t)$ with *significant* harmonic distortion. For these reasons, natural-sampling PWM systems have been considered for use in high-fidelity and/or high-efficiency audio amplifier systems [1], [4]. Since nowadays many audio sources are not continuous-time signals $x(t)$ but rather sequences of digitized samples of $x(t)$, a completely digital PWM audio amplifier is an intriguing possibility. Note that it is easy to construct a trailing-edge PWM signal pulse from a digital $B$-bit sample value which we shall assume is an unsigned integer in the range $[0, 2^B - 1]$ The modulator output goes high at the start of the pulse interval, and a digital downcounter clocked at a frequency $2^B f_c$ is used to count down the input to 0 – at which time the modulator output goes low. Unfortunately, since digital audio signals are *uniformly spaced* samples, the output signal from such a *digital pulse width modulator* is an uniform-sampling PWM signal instead of the desired natural-sampling PWM signal. The solution to this problem is to use the simple digital pulse width modulator but to replace each uniform sample value with the corresponding digitally computed natural sample value. The pulse widths are thus proportional to the natural sample values, and the digital modulator output is therefore a natural sampling PWM signal. Note that these computed natural sample values are uniformly spaced in time.

---
[1] It was shown in [2] that the intersection point is unique if $f_c > \pi f_s$ where $f_s$ is the highest frequency in $x(t)$.

## C. Algorithm I for Natural Sample Value Computation

As described in [1], [4], many different methods have been proposed for the computation of natural sample values from uniform sample values (also called natural-sampling conversion.) Many of these are too computationally intensive for practical use and many are based on various *ad hoc* approximations that have only been validated empirically. In this Section, we describe Algorithm I for converting uniform samples to natural samples at the same rate. Algorithm I is based on the Natural-Sampling Theorem and the Stirling's central difference formula [5]. The Natural-Sampling Theorem is stated as follows. [2], [3], stated as follows.

*Theorem 1:* (Natural-Sampling Theorem) The natural samples of $x(t)$ are exactly the uniform samples of $\hat{x}(t + T/2)$ where

$$\hat{x}(t) = x(t) + \sum_{n=1}^{\infty} \frac{1}{(n+1)!} \left(\frac{T}{2}\right)^n \frac{d^n}{dt^n} [x(t)]^{n+1} \quad (1)$$

Since this series converges rapidly, the first few terms suffice to compute the natural sample value to a high degree of accuracy. The first four terms give

$$\hat{x}(t) \approx x(t) + \frac{1}{2}\left(\frac{T}{2}\right)\frac{d}{dt}x^2(t) + \frac{1}{6}\left(\frac{T}{2}\right)^2\frac{d^2}{dt^2}x^3(t)$$
$$+ \frac{1}{24}\left(\frac{T}{2}\right)^3\frac{d^3}{dt^3}x^4(t)$$

$$= x(t) + \left(\frac{T}{2}\right)x(t)\frac{dx(t)}{dt}$$
$$+ \frac{T^2}{4}x(t)\left(\frac{dx(t)}{dt}\right)^2 + \frac{T^2}{8}x^2(t)\frac{d^2x(t)}{dt^2}$$
$$+ \frac{T^3}{8}\frac{d^3x(t)}{dt^3} + \frac{3T^3}{16}x^2(t)\frac{dx(t)}{dt} + \frac{T^3}{48}x^2(t)\frac{d^3x(t)}{dt^3}$$

and the various derivatives can be computed from samples of $x(t)$ via standard numerical methods such as Stirling's central difference formula.

Algorithm I for natural sample value computation then can be stated as follows.
Define the $n$-th sample value of $x(t)$ as $x[n] = x((n+\frac{1}{2})T)$ and let

$$a[n] = \frac{T}{2}\frac{dx(t)}{dt}\bigg|_{t=(n+\frac{1}{2})T}$$

$$b[n] = \frac{T^2}{8}\frac{d^2x(t)}{dt^2}\bigg|_{t=(n+\frac{1}{2})T}$$

$$c[n] = \frac{T^3}{48}\frac{d^3x(t)}{dt^3}\bigg|_{t=(n+\frac{1}{2})T}$$

Then,

$$a[n] = \frac{1}{120}\Big(x[n+3] - x[n-3]\Big)$$
$$- \frac{3}{40}\Big(x[n+2] - x[n-2]\Big)$$
$$+ \frac{3}{8}\Big(x[n+1] - x[n-1]\Big)$$

$$b[n] = \frac{1}{720}\Big(x[n+3] + x[n-3]\Big)$$
$$- \frac{3}{160}\Big(x[n+2] + x[n-2]\Big)$$
$$+ \frac{3}{16}\Big(x[n+1] + x[n-1]\Big)$$
$$- \frac{49}{144}\Big(x[n]\Big)$$

$$c[n] = -\frac{1}{384}\Big(x[n+3] - x[n-3]\Big)$$
$$+ \frac{1}{48}\Big(x[n+2] - x[n-2]\Big)$$
$$- \frac{13}{384}\Big(x[n+1] - x[n-1]\Big)$$

These quantities are readily computed via finite-impulse response (FIR) digital filters, and then the (approximate) natural sample value is computed as

$$\hat{x}[n] \approx x[n]\{(1 + a[n])\left(1 + (a[n])^2\right)$$
$$+ x[n]\left((3a[n] + 1)b[n] + x[n]c[n]\right)\}$$

The actual pulse width varies from 0 to $T$ as the input value to the pulse modulator ranges from $-1$ to $+1$ and therefore the actual value of the $n$-th pulse width is $\frac{T}{2}(1 + \hat{x}[n])$.

*D. Up-sampling and Digital Interpolation*

In Section I.B above, we stated that low-pass filtering of a natural sampling PWM signal recovers the modulating signal $x(t)$ without distortion. In fact, a PWM signal contains energy in *subbands* centered around the carrier frequency $f_c$, and unless $f_c \gg f_s$, these subbands alias into the low-pass filter passband giving rise to distortion. Now, most digital audio signals are obtained by sampling $x(t)$ at rates slightly larger than the Nyquist rate $2f_s$. For example, a telephone voice signal is band-limited to 3.4 kHz and sampled at 8 kHz while high-fidelity audio signals are band-limited to 20 kHz and sampled at 44.1 kHz for recording on CDs. Direct application of the results of Section I.C generally results in too much distortion at the output of the low-pass filter demodulator. Thus, up-sampling and digital interpolation are generally used to create digital sequences at higher rates, and the digital pulse width modulator uses carrier signals at these higher rates. For example, in [1], [3], [4], the digital signals from a CD are upsampled by a factor of 8 and the digital PWM system operates with a carrier frequency of 352.8 kHz. Thus, the block diagram of a typical PWM audio amplifier is as shown in Fig. 2.

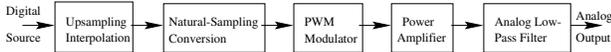

Fig. 2. Block diagram of a digital PWM audio amplifier

## II. COMBINING UP-SAMPLING, DIGITAL INTERPOLATION AND NATURAL-SAMPLING CONVERSION

In this section, we first present the Interpolation and Filtering Theorem. This theorem then will be used in deriving a new approach to combine the first two functional blocks in Fig. 2. The new algorithm, Algorithm II, will be initially described in an intuitive way. Later, we will show that the implementation structure uses the poly-phase components of a digital interpolator and digital differentiators, followed by a non-linear stage.

Now, we examine the system in Fig. 2 with some assumptions. First, as mentioned in the previous section, suppose that the original digital source is sampled at rate $f_1 = 44.1$ kHz. This signal is then up-sampled to rate $f_2 = 8f_1 = 352.8$ kHz. Second, we will imitate a 65-tap FIR filter using Hamming window for digital interpolation. This filter approximates the ideal digital interpolation filter, which is an ideal low-pass filter with cut-off frequency $\pi/8$ and magnitude in the pass-band 8. Before going on, we present the following theorem, which is one point of departure of the new approach.

*Theorem 2:* (Interpolation and Filtering Theorem) Given a finite-length input sequence of length $(2k+1)$ $\{x[n]\}_{-k}^{k}$ and a sampling period $T$, two below methods yield the same output sample at time $\tau$ where $-kT \leq \tau \leq kT$.

*Method 1.* Fitting a curve through $(2k+1)$ points of $x[n]$ using the formula

$$x(t) = \sum_{n=-k}^{k} x[n] f(t-nT) \qquad (2)$$

then sampling $x(t)$ at time $t = \tau$. Here we call $f(t)$ the interpolation function.

*Method 2.* Calculating $x(\tau)$ using convolution

$$x(\tau) = \sum_{n=-k}^{k} x[n] h[-n] \qquad (3)$$

where $\{h[n]\}_{-k}^{k}$ is given by

$$h[n] = f(t+\tau)|_{t=nT} \ , \ n = -k \div k \qquad (4)$$

The proof of this theorem is simple and will be given in the full paper.

Based on this theorem, what is done by the first two blocks in Fig. 2, Upsampling-Interpolation and Natural-Sampling Conversion, is equivalent to the following. First a curve is fit through 9 input samples at rate $T_1 = 1/f_1$ of each window. The interpolation function here is a Hamming-windowed sinc function. Then this curve is sampled at rate $T_2 = 1/f_2 = T_1/8$ to produce the signal at rate $f_2 = 8f_1$. Now, if we use Algorithm I for the natural-sampling conversion block, we have to fit another curve through every 7 rate-$f_2$ points using the Stirling's central difference formula. This block then calculates the derivatives of the curve and evaluate them at the central point. The signal and its derivatives at rate $f_2$ are finally used to compute the natural sample.

Thus, we see that separate implementation of the first two blocks in Fig. 2 is equal to doing curve fitting twice. Obviously, there are some simple ways to improve the implementation. First, we observe that digital interpolation and natural-sampling conversion both employ FIR filters. Thus, the filters can be combined using convolution. Second, we can use poly-phase implementation, either for the digital interpolation filter separately or for the combined filters. These changes, however, do not enhance the accuracy of the natural-sampling conversion.

Now, if instead, we use the very first curve after digital interpolation to calculate the derivatives, we can reduce the implementation cost and produce more accurate natural samples. Thus, the ideas of the new approach can be briefly described as follows.

*Algorithm II for Natural Sample Value Computation*

1) Fit a curve through 9 input rate-$f_1$ uniform samples $x[n-4]$, $x[n-3]$, $x[n-2]$, $x[n-1]$, $x[n]$, $x[n+1]$, $x[n+2]$, $x[n+3]$, $x[n+4]$ (Sample $x[n]$ is at time $t = 0$). The interpolation function is a Hamming-windowed sinc function

$$\begin{aligned} f(t) &= \left[ sinc\left(\frac{t}{T_1}\right) \right] \\ &\times \left[ 0.54 - 0.46 \cos\left(\frac{2\pi(t+32T_1)}{64T_1}\right) \right] \\ &\times \left[ u(t+32T_1) - u(t-32T_1) \right] \end{aligned} \qquad (5)$$

where $sinc(t) = \frac{\sin(\pi t)}{\pi t}$ and $u(t)$ is the unit step function.

The equation of the curve is then given by

$$x(t) = \sum_{i=-k}^{k} x[n+i] f(t + iT_1) \quad (6)$$

2) Compute the order-1 to order-$l$ derivatives of $x(t)$

$$x^{(l)}(t) = \sum_{i=-k}^{k} x[n+i] f^{(l)}(t + iT_1) \quad (7)$$

Note that $x(t)$ is piecewise continuous within $[-4T_1, 4T_1]$ and the discontinuities are at multiples of $T_1$.

3) Sample $x(t)$ and its derivatives at 8 points around 0: $-7T_2/2$, $-5T_2/2$, $-3T_2/2$, $-T_2/2$, $T_2/2$, $3T_2/2$, $5T_2/2$, and $7T_2/2$ ($T_2 = T_1/8$.)
4) Use the values from *Step* 3 in the Natural-Sampling Theorem (1) to calculate 8 consecutive natural samples at rate $f_2 = 8f_1$.
5) Shift the window one input sample forward and go back to *Step* 1.

From (6) and (7), we can see that each sample of the signal and its derivatives is just a linear combination of the 9 input samples. Thus, the rate-$f_2$ samples of the signal and its derivatives can be calculated using length-9 FIR filters. Using Theorem 2, we can also prove that the coefficients of digital interpolation filters for each of 8 samples are just the poly-phase components of a filter $\{h_0[n]\}$ where $h_0[n] = f(t)|_{t=nT_2+T_2/2}$, $n = -32 \div 31$. Similarly, the order-$l$ digital differentiators are the poly-phase components of the filter $\{h_l[n]\}$ where $h_l[n] = f^{(l)}(t)|_{t=nT_2+T_2/2}$, $n = -32 \div 31$. Thus to calculate the coefficients, we just have to evaluate $f(t)$ and its derivatives at these time instants.

In the algorithm described above, we have used simple digital differentiators, which are just rectangular-windowed derivatives of the Hamming-windowed sinc function. However, the algorithm can be easily generalized for general digital interpolation filters and digital differentiators. Note that although we are not using Stirling's formula in Algorithm II, it is still an eligible candidate for these differentiators.

In what follows, we discuss the improvement of Algorithm II over Algorithm I used together with an Upsampling and Digital Interpolation Block. Note that, as mentioned previously, curve fitting is used only as an interpretation of the filtering process. The linear parts of the real systems will be implemented using FIR filters. If the Natural-Sampling Conversion is implemented separately after the Upsampling and Digital Interpolation, we will have to compute the derivatives of $x(t)$ based on the interpolated uniform samples, which come with distortion after a non-ideal digital interpolator. Instead, if we use Algorithm II, we can apply the digital differentiators right to the low-rate uniform samples, thus enhance the accuracy of the natural samples. (The digital interpolator, whereas, can be kept the same.) Furthermore, this approach allows the use of poly-phase implementation of both the digital interpolator and digital differentiators, therefore reduce the implementation cost of the whole system.

Let $K$ denote the number of terms we take in the Natural-Sampling Theorem. Also, let $x[m]$ be the rate-$f_2$ samples of the signal $x(t)$, and $a[m]$, $b[m]$, $c[m]$ be the samples of $x'(t)/2$, $x''(t)/4$ and $x'''(t)/8$, respectively (The derivatives have been scaled down to simplify the implementation.) From the Natural-Sampling Theorem, the natural sample is given by

$$\begin{aligned}
K = 1: z[m] &= s[m] \\
K = 2: z[m] &= s[m](1 + a[m]) \\
K = 3: z[m] &= s[m]\left(1 + a[m] + a[m]^2 + s[m]b[m]\right) \\
K = 4: z[m] &= s[m]\{(1 + a[m])\left(1 + a[m]^2\right) \\
&+ s[m]\left((3a[m] + 1)b[m] + s[m]c[m]\right)\}
\end{aligned} \quad (8)$$

Note that when $K = 1$, the output is just the uniform-sampling signal at rate $f_2$.

We could take more terms in the Natural-Sampling Theorem to improve the accuracy of the conversion.

A poly-phase implementation of the linear part for $K = 4$ is given in Fig. 3. The non-linear part is described in Fig. 4.

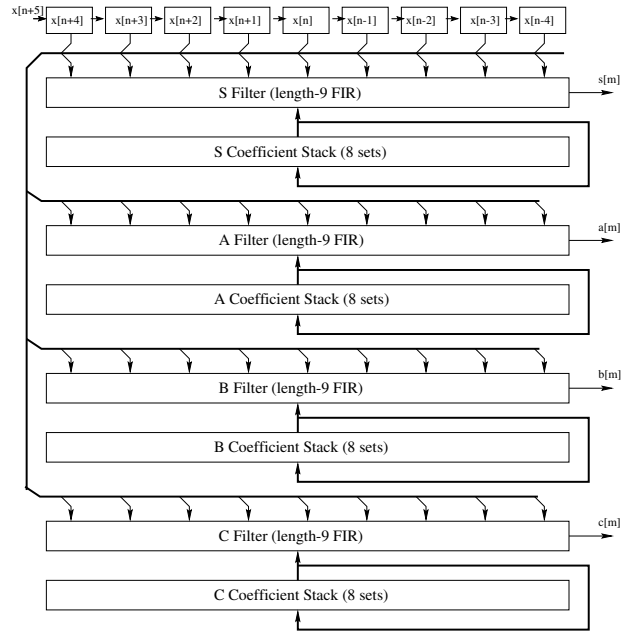

Fig. 3. Implementation of the linear part

### III. SIMULATION RESULTS

Some simulation results for $K = 1, 2, 3$ and $4$ are given in Fig. 5. The digital interpolation filter and the digital differentiators employ the Hamming-windowed sinc function and its derivatives, as described in Section II. The modulating signal is a 6.6 kHz sinusoid. The demodulator is an ideal analog low-pass filter with cut-off frequency 20 kHz. It can be seen that the second-order and third-order harmonics (13.2 kHz and 19.8 kHz, respectively) decrease as K increases. More specifically, it can be observed that the second and the third harmonics alternately decay. Further simulation results, analysis and comparison will be presented in detail at the conference.

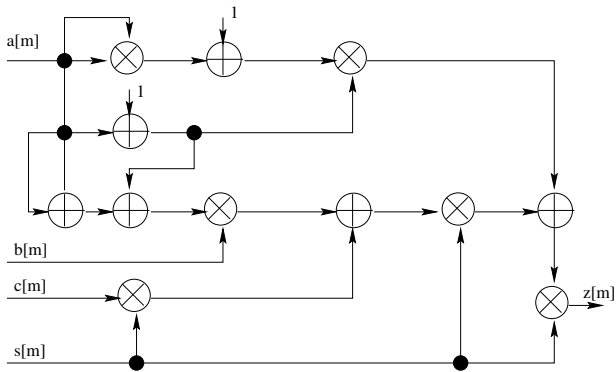

Fig. 4. Implementation of the non-linear part

## IV. Concluding remarks

In this paper, we have proposed a new approach to implement the upsampling, digital interpolation and natural-sampling conversion for PWM audio amplifiers. This approach employs poly-phase implementation of digital interpolation filter and digital differentiators to compute the higher-rate natural samples directly from the original low-rate uniform-samples. Therefore, it can reduce the implementation cost and enhance the accuracy for natural-sampling conversion. This method can also easily be expanded to accomodate more terms in the Natural-Sampling Theorem, thus producing better natural samples at the expense of more complicated implementation.

If the number of terms in the Natural-Sampling Theorem is fixed, choosing the right digital interpolation filter and digital differentiators are the key to have good natural samples.

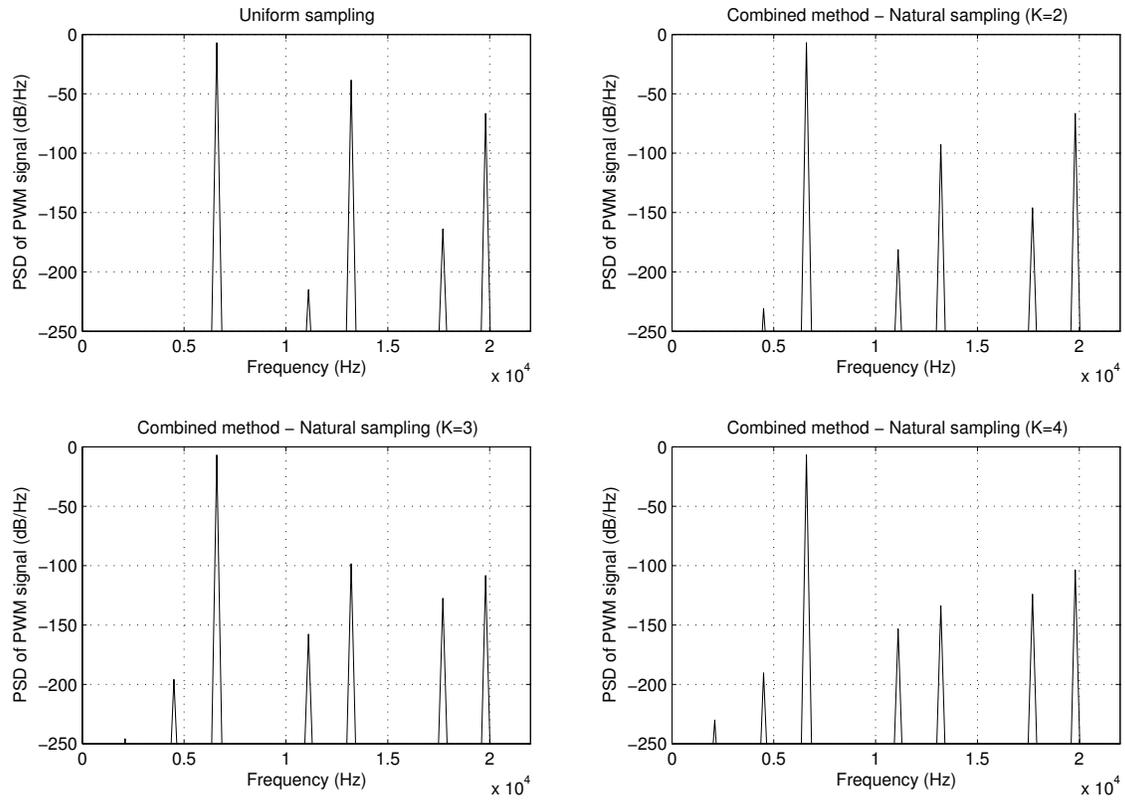

Fig. 5. Simulation results for K=1, 2, 3 and 4. The digital interpolation filter and the digital differentiators employ the Hamming-windowed sinc function and its derivatives